# Deep Learning Model for Detecting Abnormal Corn Kernels

Suwannee Adsavakulchai* | Mawin Prommasaeng

*Abstract* - This research aims to detect the physical characteristics of corn kernels and analyze images using a deep learning model. The data analysis based on the CRISP-DM framework which consists of six steps, business understanding, data understanding, data preparation, modelling, evaluation, and deployment. The business goal reduces the cost of the separation of abnormal corn kernels. The dataset comprises 1,800 images of corn kernels and divided equally between normal and abnormal corn kernels. The dataset was divided into three subsets: 1,000 images for training the deep learning model, 600 images for validation and 200 images for testing. The tools for analysis in this research are Jupyter Lab, Python, TensorFlow Keras, and Convolutional Neural Networks. The results revealed that the deep learning model achieved the accuracy rate of 99% in differentiating between normal and abnormal corn kernel images that is a highly effective model in this context.

## 1. Introduction

Corn is one of the raw materials that uses in many manufacturing processes, it is very important to assess its quality in terms of physical and chemical characteristics. Physical quality checks the moisture content, fungal inspection, counterfeit identification and seed assessment [1]. The examination of seed detects the cracked, split, or damaged seeds, which currently relies on human labor, including visual observation, counting, comparison with reference samples, and calculations. However, this manual approach has limitations due to human variability and expertise discrepancies, leading to slow, inaccurate, non-standardized assessments, and high costs [2].

Advancements technology both in hardware and software led to the integration of various technologies to replace human labor. Data is transformed into digital format to enhance business processes. Artificial Intelligence increase the analysis and prediction to automate workflows, speed, accuracy, and efficiency across various stages. In term of the corn quality inspection, digital image processing and AI are used to help the human eye and brain's analytical capabilities. Deep Learning is a subset of AI that empowers better decision-making, resulting in more precise predictions and a competitive in the business arena [3,4].

Deep learning is to classify and inspect the physical quality of corn kernels. By studying normal and abnormal corn kernel characteristics and converting them into digital images, an Artificial Neural Network (ANN) is trained using deep learning techniques. The research revealed the effectiveness of the model in accurately classifying normal and abnormal corn kernels for businesses decisions and a competitive advantage in the market [5,6].

The purposes of this research are 1. to study the physical characteristics of normal and abnormal corn kernels 2. to develop a deep learning model for analyzing abnormality in corn kernel images and 3. to develop the web application for analyzing abnormality in corn kernel images.

## 2. Research Methodology

Research framework for analyzing abnormality in corn kernel images is shown in Figure 1

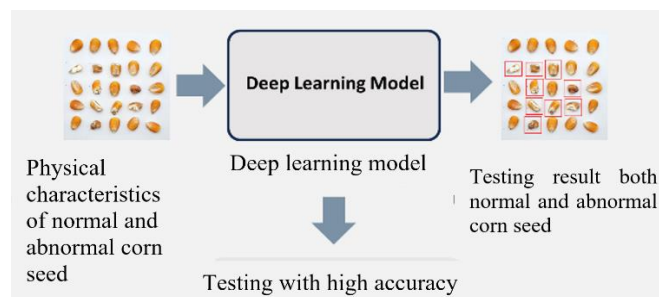

Figure 1 Research framework

*Corresponding author Suwannee Adsavakulchai, School of Engineering, University of the Thai of Commerce, Bangkok, THAILAND 10400 (e-mail: suwannee_ads@utcc.ac.th)

From Figure 1, input/output image data of normal and abnormal corn kernels to develop a model for classifying corn kernels. Deep learning research for detecting abnormalities in corn kernels follows the CRISP-DM process [7] that consisting of 6 main steps:

*I. Problem Understanding*
Study and understand the physical quality inspection problem related to abnormality detection in corn kernels. This step establishes the foundation for the research.

*II. Data Understanding*
It is a critical step in data analysis process, including the analysis of the charnormal and abnormal corn kernels. This step involves gaining insights into the dataset and its characteristics, which is essential for building an effective model. Here are some key aspects of data understanding in the context of this research: A proper understanding of the data is crucial for building an effective model.

*III. Data Preparation*
Prepare image data of both normal and abnormal corn kernels to be suitable for input into the model.

*IV. Model Development*
Create and develop a model for classifying abnormal corn kernels. Utilize various tools such as Jupyter Lab, Python, and TensorFlow Keras.

*V. Evaluation*
Assess the model's capability in classifying abnormal corn kernels and validate the accuracy of the developed model.

*VI. Deployment*
Deploy the developed model into real-world applications for practical quality assessment of corn kernels.

## 3. Results

*I. Result of Problem Understanding*
The current problem lies in the manual inspection of abnormal corn kernels, where human workers visually examine and sort them. This process is time-consuming and labor-intensive, resulting in increased labor costs due to the rising cost of labor. To address these challenges, there require enhancing the efficiency of abnormal corn kernel inspection by technology. This will not only accelerate the inspection process but also reduce work time. The goal is to replace manual labor, improve competitiveness, and foster consumer confidence.

*II. Result of Understanding the Data*
The separation of abnormal corn kernels from normal ones requires a comprehensive understanding of the physical characteristics of both types.

- Physical Characteristics of Normal Corn Kernels:
  Normal corn kernels are fully developed with a size ranging from 0.5 to 0.8 cm. The surface of the kernel is smooth, glossy, and firm. The color is vary between yellow and slightly orange. The inner part is covered by the husk known as "endosperm," which is white or pale yellow. The tip of the kernel is white. Normal kernels are not wrinkled, cracked, or damaged by insects or animals. They emit a pleasant odor and do not have a musty or off-putting smell. There should be no presence of green, black, or any other unusual colors that differ from the typical color of corn kernels. The physical characteristics of normal corn kernels are depicted in Figure 2.

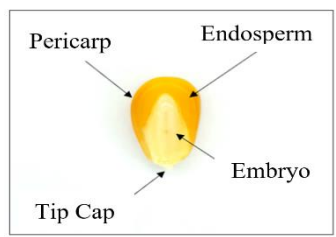

Figure 2    The physical characteristics of normal corn kernels

From Figure 2, the characteristics of normal corn kernels in comparison to the abnormal ones. It provides guidance for accurately identifying the different types based on their physical attributes.

- Physical Characteristics of Abnormal Corn Kernels:
  Abnormal corn kernels cracked or incomplete, wrinkled, distorted, show signs of being damaged by insects or animal bites, fungal growth attached to the kernel. They have green, black, or other colors that differ from the normal color of corn kernels. The physical characteristics of abnormal corn kernels are distinct from those of normal kernels are depicted in Figure 3.

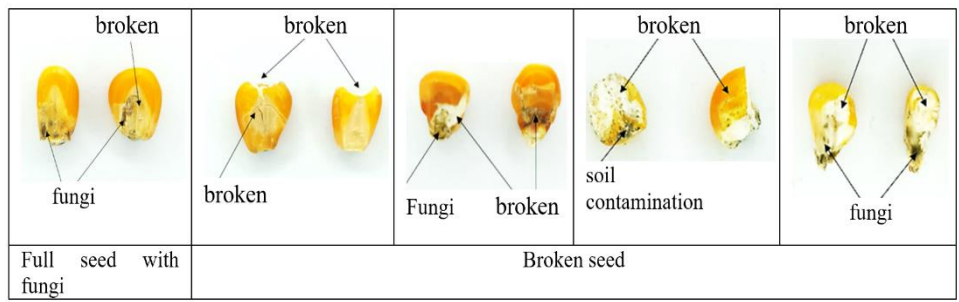

Figure 3 The physical characteristics of abnormal corn kernels

From Figure 3, the distinguishing features of abnormal corn kernels, aiding in their accurate identification and differentiation from normal kernels based on their physical attributes.

## III. Data Preparation Results

The corn kernels are divided into two groups: normal corn kernels and abnormal corn kernels. The dataset is prepared by capturing images of corn kernels with a size of 250x250 pixels. A total of 1,800 images are included in the dataset, comprising 900 images of normal corn kernels and 900 images of abnormal corn kernels. These images are utilized to develop a model, with 1,000 images for training, 600 images for validation, and 200 images for testing presented in Table 1. Examples of the image dataset showed in Figures 4 and 5.

Table 1 Images to develop a model.

| Train Set | | Validate Set | | Test Set | |
|---|---|---|---|---|---|
| Normal | Abnormal | Normal | Abnormal | Normal | Abnormal |
| 500 | 500 | 300 | 300 | 100 | 100 |

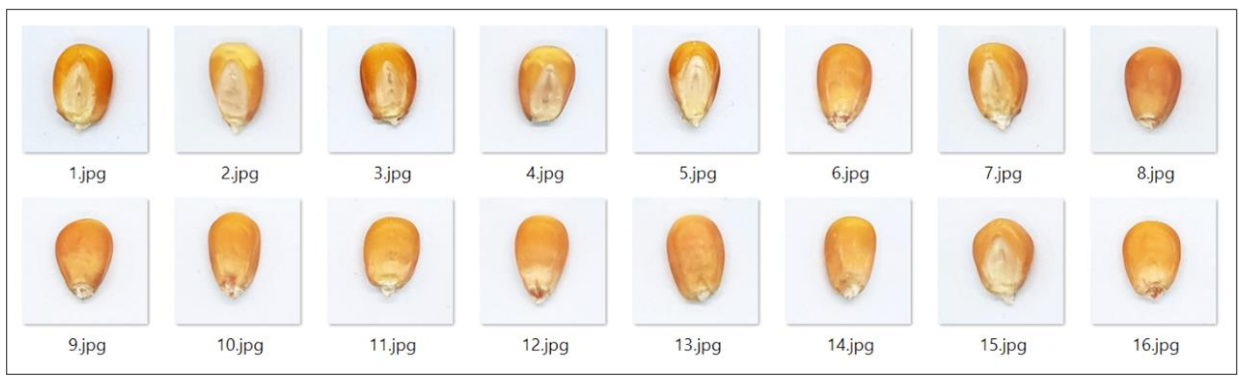

Figures 4 Sample of normal corn kernels dataset

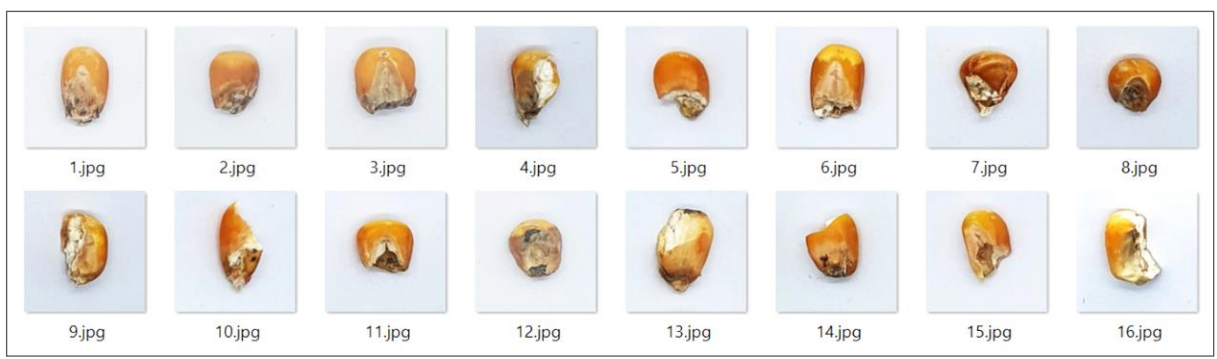

Figures 5 Sample of abnormal corn kernels dataset

From Figure 4 and 5, the data preparation process is for training, validating, and testing the model using the collected images of normal and abnormal corn kernels, fostering the development of an effective classification model.

*IV. Model Development*

The development of the model is carried out using Jupiter Lab, Python, and the TensorFlow Keras library [1,8]. The dataset of corn kernel images is split into two groups for training the model. The first group is the "Train set," which includes images of normal corn kernels, and the second group is images of abnormal corn kernels. The images are loaded and processed using Python libraries like OpenCV and NumPy within the Jupiter Lab environment.

The model is constructed using the TensorFlow Keras framework. a Convolutional Neural Network (CNN) architecture is chosen for the model [1]. CNNs are deep learning models designed for image classification tasks. The training process involves using the Train Set data from both groups. Important parameters such as the number of epochs, learning rate, and batch size are configured [1] to optimize the model's performance.

By utilizing these tools and techniques, the CNN model is developed and trained to classify the corn kernel images into normal and abnormal categories, leveraging the power of deep learning for accurate classification.

After completing the training process, the model is validated using the Validation Set data to ensure its accuracy in classifying images correctly. Once the model achieves a satisfactory level of accuracy, the process of creating the final model is initiated using TensorFlow Keras and the Convolutional Neural Network (CNN) architecture [1]. This model process of testing the model's capability to accurately identify abnormal corn kernels using unseen data. The steps involved in creating and testing the final model for identifying abnormal corn kernels are demonstrated in Figure 6.

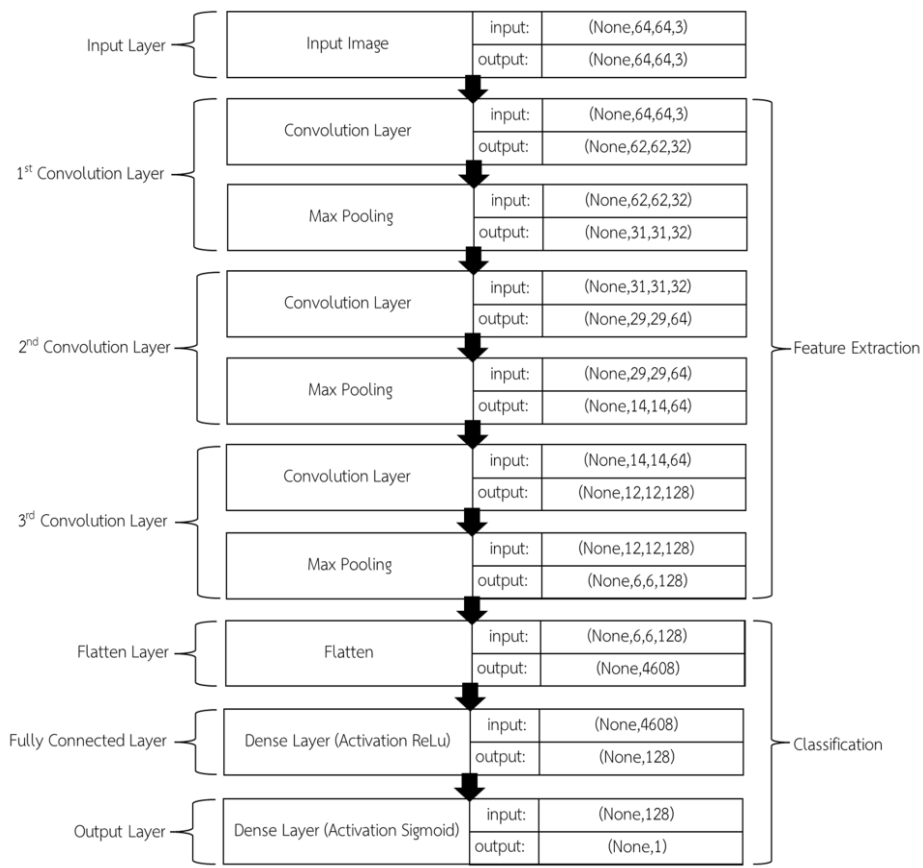

Figure 6 CNN architecture

From Figure 6, the structure of the final model, which employs the CNN architecture, illustrates the arrangement of layers and connections within the model, which plays a crucial role in making accurate predictions based on input images. After successful validation and fine-tuning, the final model is ready to be deployed for real-world applications, demonstrating its ability to effectively identify abnormal corn kernels using the power of deep learning and CNNs.

*V. Evaluation Results*

The results of training the model over 1,000 epochs show that as the training progresses, the loss decreases, leading to higher accuracy. After training, the model achieved a loss value of 0.0475, as shown in Figure 7, and an accuracy rate of 0.9900, which is equivalent to 99%, as displayed in Figure 8. The graphs illustrate the accuracy during training and validation.

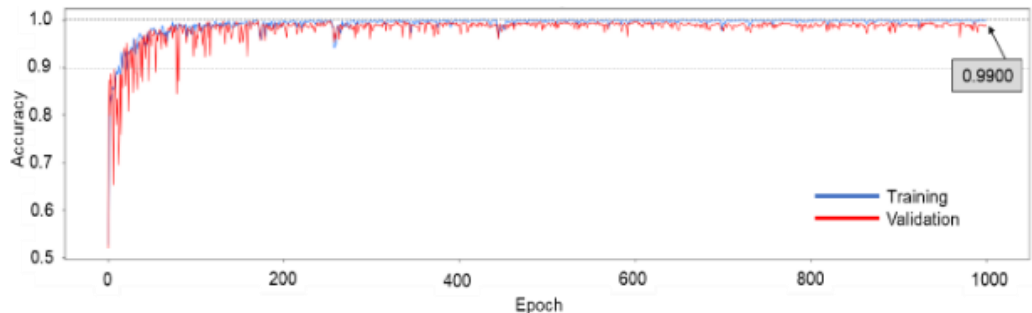

Figure 7 The accuracy of the model during training and validation.

From Figure 7, at the beginning of training, the accuracy was low, but as training continued, the accuracy steadily increased.

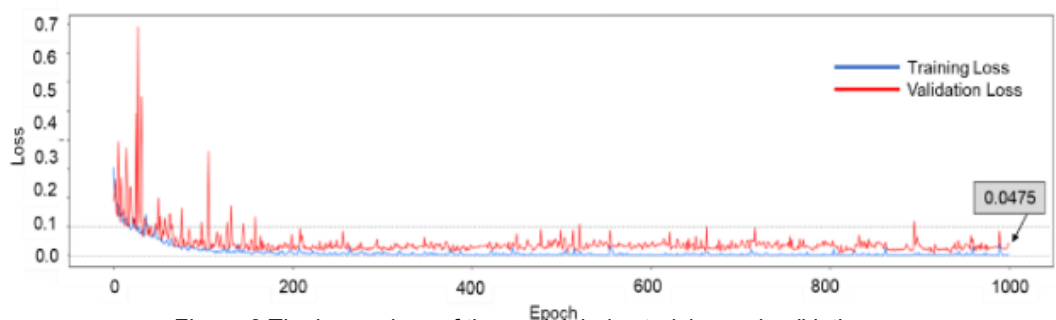

Figure 8 The loss values of the model during training and validation

From Figure 8, at the beginning of training, the loss was high, but as training proceeded, the loss consistently decreased.

The classification is determined based on the output of the Sigmoid function. When the output is less than or equal to 0.5, it is classified as "Abnormal" corn seeds. Conversely, when the output is greater than 0.5, it is classified as "Normal" corn seeds. The testing of the model for predicting "Normal" and "Abnormal" corn. The prediction values are presented in Table 2.

Table 2 The accuracy of the testing results

| Seed# | actual | calculation | predict | Seed# | actual | calculation | predict |
|---|---|---|---|---|---|---|---|
| 1.jpg | Abnormal | 0.000 | Abnormal | 11.jpg | Normal | 1.000 | Normal |
| 2.jpg | Abnormal | 0.000 | Abnormal | 12.jpg | Normal | 1.000 | Normal |
| 3.jpg | Abnormal | 0.000 | Abnormal | 13.jpg | Normal | 1.000 | Normal |
| 4.jpg | Abnormal | 0.000 | Abnormal | 14.jpg | Normal | 1.000 | Normal |
| 5.jpg | Abnormal | 0.000 | Abnormal | 15.jpg | Normal | 1.000 | Normal |
| 6.jpg | Abnormal | 0.000 | Abnormal | 16.jpg | Normal | 1.000 | Normal |
| 7.jpg | Abnormal | 0.000 | Abnormal | 17.jpg | Normal | 1.000 | Normal |
| 8.jpg | Abnormal | 0.000 | Abnormal | 18.jpg | Normal | 1.000 | Normal |
| 9.jpg | Abnormal | 0.000 | Abnormal | 19.jpg | Normal | 1.000 | Normal |
| 10.jpg | Abnormal | 0.000 | Abnormal | 20.jpg | Normal | 1.000 | Normal |

From Table 2, the testing results for the model's predictions in identifying "Normal" and "Abnormal" corn seeds indicate a remarkable accuracy rate of 100%. This means that the model correctly classified all instances of "Normal" and "Abnormal" corn seeds during testing. The evaluation of the model's accuracy yielded an accuracy rate of 0.990 or 99%, which is considered a very high and acceptable level of accuracy.

*VI. Deployment*

The model was successfully integrated into an application software designed specifically for its use, as depicted in Figure 9.

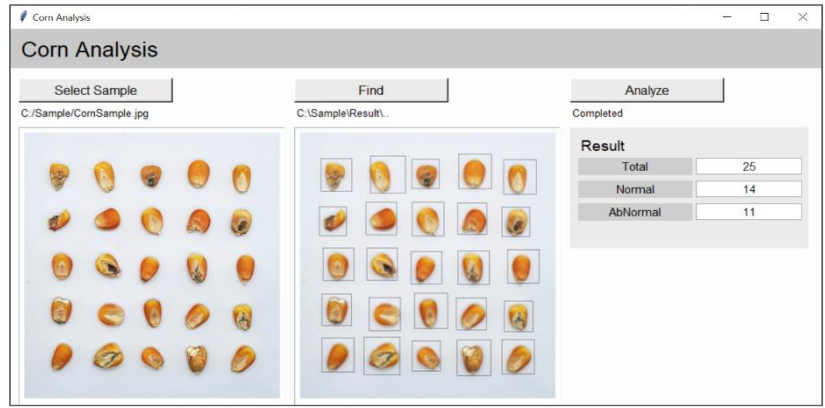

Figure 9 The application of the model for corn seed analysis.

From Figure 9, the application, the photographs of corn seeds with different characteristics were inputted. The program automatically detected the positions of the corn seeds within the images. Subsequently, the model was invoked to analyze these seeds. This application facilitates the practical implementation of the model for corn seed analysis. The result of the model for corn seed analysis showed in Figure 10.

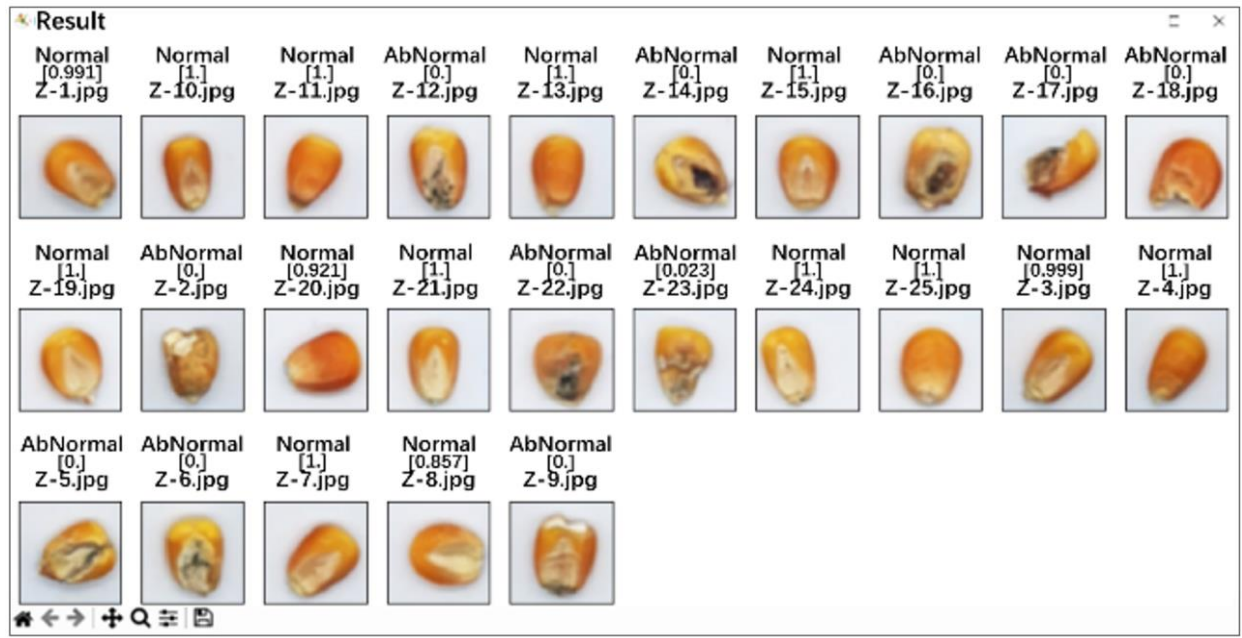

Figure 10 The result of the model for corn seed analysis

Figure 10, the analysis results are depicted a total of 25 corn seeds were analyzed. Out of these, 14 seeds were correctly identified as "Normal," while 11 seeds were correctly identified as "Abnormal." The analysis outcomes are presented in Table 3.

Table 3 The accuracy of the analysis results

| Seed# | actual | calculation | predict | Seed# | actual | calculation | predict |
|---|---|---|---|---|---|---|---|
| Z-1.jpg | Normal | 0.991 | Normal | Z-14.jpg | Abnormal | 0.000 | Abnormal |
| Z-2.jpg | Abnormal | 0.000 | Abnormal | Z-15.jpg | Normal | 1.000 | Normal |
| Z-3.jpg | Normal | 0.999 | Normal | Z-16.jpg | Abnormal | 0.000 | Abnormal |
| Z-4.jpg | Normal | 1.000 | Normal | Z-17.jpg | Abnormal | 0.000 | Abnormal |
| Z-5.jpg | Abnormal | 0.000 | Abnormal | Z-18.jpg | Abnormal | 0.000 | Abnormal |
| Z-6.jpg | Abnormal | 0.000 | Abnormal | Z-19.jpg | Normal | 1.000 | Normal |
| Z-7.jpg | Normal | 1.000 | Normal | Z-20.jpg | Normal | 0.921 | Normal |
| Z-8.jpg | Normal | 0.857 | Normal | Z-21.jpg | Normal | 1.000 | Normal |
| Z-9.jpg | Abnormal | 0.000 | Abnormal | Z-22.jpg | Abnormal | 0.000 | Abnormal |
| Z-10.jpg | Normal | 1.000 | Normal | Z-23.jpg | Abnormal | 0.023 | Abnormal |
| Z-11.jpg | Normal | 1.000 | Normal | Z-24.jpg | Normal | 1.000 | Normal |
| Z-12.jpg | Abnormal | 0.000 | Abnormal | Z-25.jpg | Normal | 1.000 | Normal |
| Z-13.jpg | Normal | 1.000 | Normal | | | | |

From Table 3, the model's testing results for classifying "Normal" and "Abnormal" corn seeds, it achieved an accuracy rate of 99%, indicating high performance in distinguishing between these categories. The prediction accuracy for the analyzed seeds was 100% that the model provided completely accurate results.

## 4. Discussion

The physical quality assessment, corn seeds can be categorized into two groups: "Normal" and "Abnormal." Each group displayed the physical characteristics that are similar, as illustrated in Figures 4 and 5. In this research, Deep Learning techniques were employed to build a model for the purpose of corn seed inspection. The model was trained using images of both "Normal" and "Abnormal" corn seeds. Tools such as Jupyter Lab, Python, TensorFlow Keras, and Convolutional Neural Networks (CNN) were utilized.

The developed model had a structure consisting of three convolutional layers and used the Sigmoid function for activation, allowing for clear classification into two distinct classes: "Normal" and "Abnormal." When applying the model, it achieved an impressive accuracy rate of 99% after 1,000 training epochs. It's worth noting that the research conducted by Henry O. Velesaca et al. [10], which also employed Deep Learning principles using Mask R-CNN for corn seed classification, achieved accuracy rates of 96.0% for "Good" seeds, 69.5% for "Defective" seeds, and 28.6% for "Impurities."

These results highlight the effectiveness of the Deep Learning approach in accurately classifying corn seeds into "Normal" and "Abnormal" categories, surpassing the performance of previous methods

## 5. Conclusion

The research focused on assessing the physical quality of corn seeds, categorizing them into two groups: "Normal" and "Abnormal." Deep Learning techniques, specifically Convolutional Neural Networks (CNN), were employed to develop a model for seed classification. The model was trained using images of both "Normal" and "Abnormal" corn seeds. The resulting model demonstrated the accuracy rate of 99% after 1,000 training epochs.

The model was integrated into an application software designed for practical use. This application allowed the analysis of corn seeds, achieving 100% accuracy in classifying them into "Normal" and "Abnormal" categories, as shown in Table 3. The model proved to be beneficial in aiding seed inspectors, enabling quick decision-making, reducing work time, and preventing losses due to inspection errors. It also instilled confidence in customers.

It can be concluded that the research has demonstrated the effectiveness of deep learning techniques in accurately classifying corn seeds, offering practical benefits and laying the groundwork for further advancements in the field of seed quality assessment.

Suwannee Adsavakulchai, et al.: Deep Learning Model for Detecting Abnormal Corn Kernels

## RECOMMENDATIONS

- Image preparation, the images used to create the dataset are standardized and of high quality to minimize variability and increase model accuracy.
- Data augmentation, to consider augmenting the training data with additional images or applying transformations to improve the model's accuracy in predicting real-world scenarios.
- Equipment calibration when using the model in an application, ensure that the imaging equipment is appropriately calibrated for factors such as lighting conditions, focal length, and image parameters. Changes in equipment may require retraining the model with new data.
- Thoroughly validate the model's performance and accuracy before deploying it in practical applications. Acceptable validation scores should be achieved before implementing the model.

On-going research could focus on developing deep learning-based machinery for corn seed sorting and quality control. The knowledge gained from this study could be extended to other types of seeds or adapted for various applications. The use of deep learning models for image analysis and classification has immense potential in improving efficiency and accuracy in various industries.

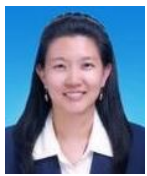
**Suwannee Adsavakulchai** received the M.S. degree from Assumption University, Bangkok in 1992 in Computer Information System and Doctor of Technical Science degree from Asian Institute of Technology in 2000. She is currently Director of Master Program in Computer Engineering and Finance Technology, School of Engineering, University of the Thai Chamber of Commerce. Her research interests include machine learning, database system, software engineering.

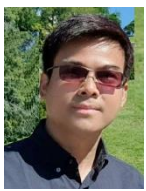
**Mawin Prommasaeng** received the B.Eng. degree from Ubon Ratchatanee University in 2001 and the M.S. degree from the University of the Thai Chamber of Commerce in 2022. His research interests include machine learning.